\documentclass{article}
\usepackage{silence}

\usepackage[final]{neurips_2024}

\usepackage[utf8]{inputenc} 
\usepackage[T1]{fontenc}    
\usepackage{hyperref}       
\usepackage{url}            
\usepackage{booktabs}       
\usepackage{amsfonts}       
\usepackage{nicefrac}       
\usepackage{microtype}      
\usepackage{xcolor}         
\usepackage{booktabs}
\usepackage{amsmath}
\usepackage{graphicx}
\usepackage{}

\title{GNN For Muon Particle Momentum estimation}

\author{
Vishak K Bhat \\
IIT Dhanbad, India \\
\texttt{vishak.bhat5@gmail.com}
\And
Eric A. F. Reinhardt \\
Department of Physics and Astronomy \\
The University of Alabama, USA \\
\texttt{eareinhardt@crimson.ua.edu}
\And
Sergei Gleyzer \\
Department of Physics and Astronomy \\
The University of Alabama, USA \\
\texttt{sgleyzer@ua.edu}
}

\begin{document}

\maketitle

\begin{abstract}
  Due to a high rate of overall data generation relative to data generation of interest, 
the CMS experiment at the Large Hadron Collider uses a combination of hardware- and software-based triggers to 
select data for capture. Accurate momentum calculation is crucial for improving the
efficiency of the CMS trigger systems, enabling better classification of low- and high-
momentum particles and reducing false triggers. This paper explores the use of Graph 
Neural Networks (GNNs) for the momentum estimation task. We present two graph construction methods and apply a GNN model to leverage the inherent graph structure of the data. In this paper firstly, we show that the GNN outperforms traditional models like TabNet in terms of Mean Absolute Error (MAE), demonstrating its effectiveness in capturing complex dependencies within the data. Secondly we show that the dimension of the node feature is crucial for the efficiency of GNN.
\end{abstract}

\section{Introduction}

Many common data capture triggers used in the CMS experiment at the Large Hadron Collider rely on accurately recording the momentum of muon particles detected in the collision data. Generally, these triggers require that the muon momentum be above some minimal threshold. After particle hit data is captured by the detector and sent to the trigger stations we estimate the momentum of that particle. In this research we present a Graph Neural Network (GNN) based model to estimate the momentum of the particles.

Graph Neural Networks as seen initially in \cite{scarselli2008graph} and \cite{1555942} have emerged as powerful tools for analyzing data with inherent graph structures, such as social networks, molecular structures, and, as we explore in this paper, data from particle physics experiments. In particular, the CMS trigger stations record multiple features of high-energy Muon particles, which can be naturally represented as nodes and edges in a graph. We propose two graph construction methods and utilize a GNN to process this data, aiming to improve the accuracy of downstream tasks, such as classification. Our approach capitalizes on the message-passing mechanism inherent in GNNs to capture complex patterns within the data. We are able to observe that the GNN are able to estimate the momentum of these particles with a less error compared to Tabnet models \cite{arik2021tabnet}.

\section{Related Work}

To estimate the momentum of muon particles initally  Boosted Decision Trees (BDTs) were used as seen in \cite{Acosta_2018}.

\section{Background}
The concept of graph neural networks (GNNs) was first introduced in early research, laying the foundation for the field \cite{gori2005new}. \cite{micheli2009neural} later developed one of the earliest spatial-based graph convolutional networks (GCNs), characterized by non-recursive layers arranged in a composite architecture. Over the years, various forms of spatial-based GCNs have been proposed, each building upon and extending Micheli's initial ideas \cite{atwood2016diffusion}. On the other hand, \cite{bruna2013spectral} pioneered spectral-based GCNs, utilizing principles from spectral graph theory to perform graph convolution. Since their introduction, numerous advancements and variations of spectral-based GCNs have emerged, further refining and expanding their capabilities.

\section{Dataset and pre processing}

The high energy Muon particles are passed through the CMS trigger which has 4 stations. The dataset is created when these particles hit the stations. These stations record 7 features namely Phi, Theta, Bending Angle, Time Info, Ring Number, Front and Mask. So we have in total 28(4*7) features extracted. This data is then converted into graphs using the following methods.

\subsection{Each trigger station as a Node of the Graph:}
Each trigger station has extracted 7 features. Here each station is considered to be a node of the graph and the 7 features become the node features. A fully connected graph is created from these nodes like Figure 1.

\subsection{Each feature extracted as a Node of the Graph:}
Each trigger station has extracted 7 features. Here each feature is considered to be a node of the graph and the values of these features extracted from 4 stations become the node features. A fully connected graph is created from these nodes like Figure 2.

\begin{figure}[ht]
    \centering
    \begin{minipage}{0.45\textwidth}
        \centering
        \includegraphics[width=\linewidth]{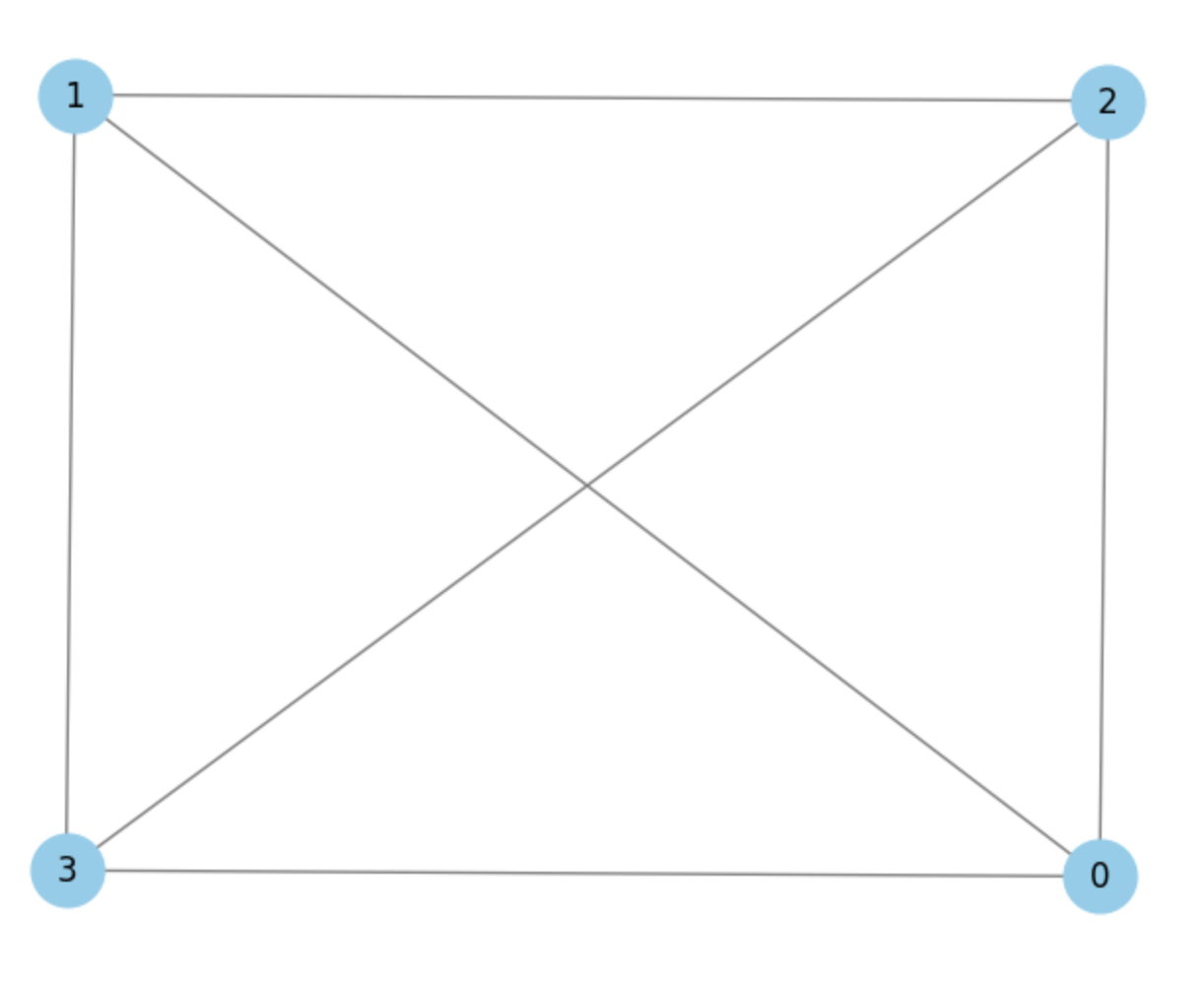}
        \caption{Each station as a Node of the Graph}
        \label{fig:image1}
    \end{minipage}\hfill
    \begin{minipage}{0.45\textwidth}
        \centering
        \includegraphics[width=\linewidth]{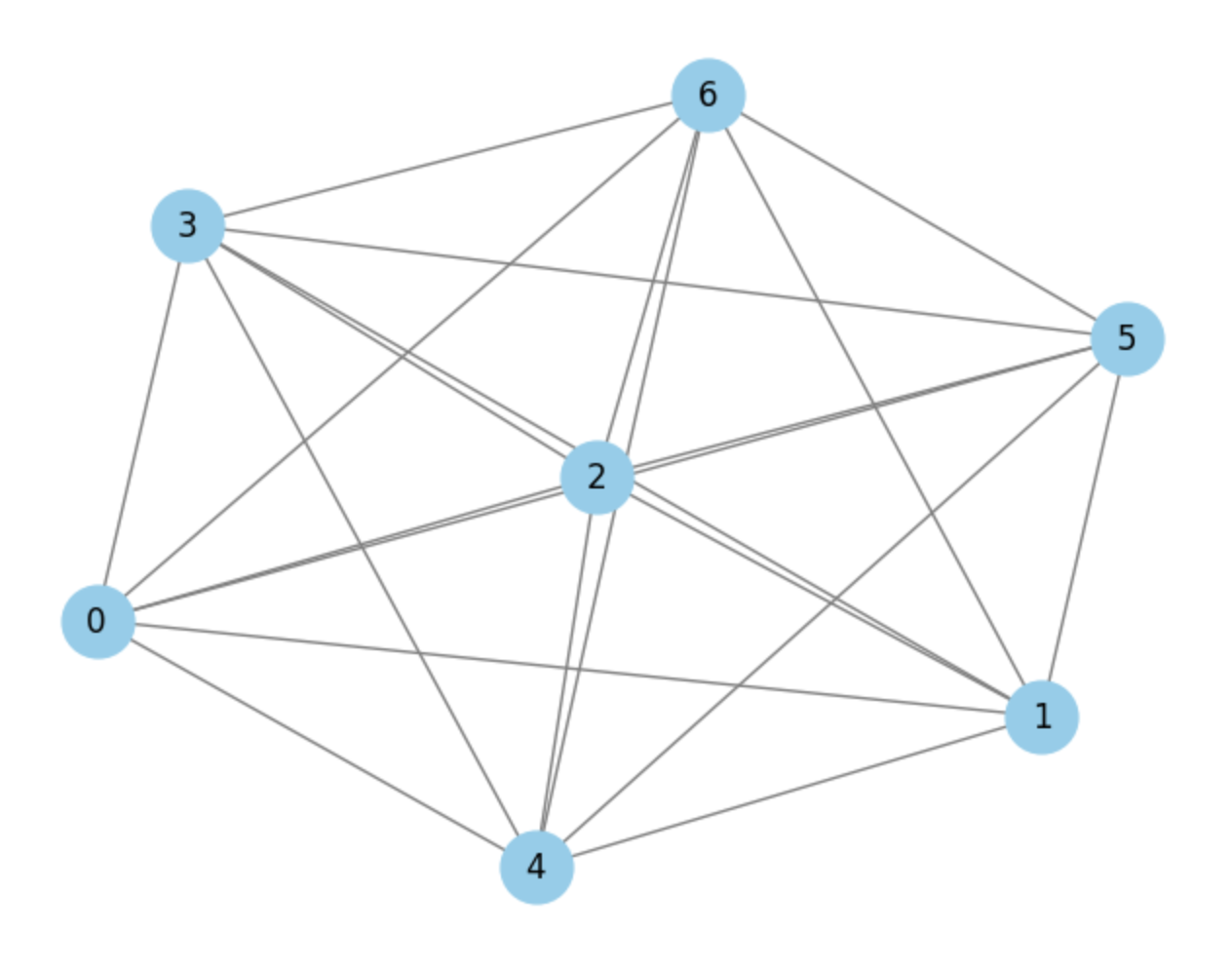}
        \caption{Each feature as a Node of the Graph}
        \label{fig:image2}
    \end{minipage}
\end{figure}

\section{Model and Training}
Message passing in GNNs enables nodes to exchange information with their neighbors. Nodes aggregate messages from adjacent nodes and update their own features based on these aggregated messages. This process helps nodes learn from their local graph structure and refine their representations iteratively. Once the representations are refined we can perform any downstream task like classification.

\subsection{Message Passing in Our Model}

In our model, the message-passing mechanism is designed as follows:

\textbf{Message Computation:}
\begin{itemize}
    \item \(\text{mlp1}\): A linear layer that computes the raw message from node \(i\) to node \(j\) by transforming the concatenated features \(x_i\) and \(x_j - x_i\):
    \[
    \text{msg}_{i \to j} = \text{ReLU}(\text{mlp1}(\text{concat}(x_i, x_j - x_i)))
    \]
    where \(\text{ReLU}\) (Rectified Linear Unit) introduces non-linearity \cite{nair2010rectified}.
    
    \item \(\text{mlp2}\): A linear layer applied to node features \(x\) to transform them from the input dimension to the output dimension:
    \[
    x = \text{ReLU}(\text{mlp2}(x))
    \]
\end{itemize}

\textbf{Weight Calculation:}
\begin{itemize}
    \item \(\text{mlp3}\) and \(\text{mlp4}\): Linear layers that compute scalar weights \(w_1\) and \(w_2\) based on the concatenated node features and messages. These weights are computed using the \(\text{Sigmoid}\) activation function \cite{rumelhart1986learning}:
    \[
    w_1 = \text{Sigmoid}(\text{mlp3}(\text{concat}(x, \text{msg}))) 
    \]
    \[
    w_2 = \text{Sigmoid}(\text{mlp4}(\text{concat}(x, \text{msg})))
    \]

    \item \(\text{mlp5}\) and \(\text{mlp6}\): Linear layers that project node features and messages into a lower-dimensional space (16 dimensions) for attention weight computation. These weights are derived using the \(\text{Tanh}\) activation function \cite{De_Ryck_2021}:
    \[
    w_1 = \text{Tanh}(\text{mlp5}(x_i)) 
    \]
    \[
    w_2 = \text{Tanh}(\text{mlp6}(\text{msg}_{i \to j}))
    \]
    
    \item \(\text{mlp7}\): Computes the final attention weight \(w\) from the product of \(w_1\) and \(w_2\), normalized using the \(\text{Softmax}\) function \cite{bridle1989training}:
    \[
    w = \text{Softmax}(\text{mlp7}(w_1 \cdot w_2), \text{edge\_index}[0])
    \]
\end{itemize}

\textbf{Aggregation and Update:}
\begin{itemize}
    \item The final node feature \(x'\) is computed as a weighted sum of the messages and the node's own features:
    \[
    x' = w_1 \cdot \text{msg}_{i \to j} + w_2 \cdot x_i
    \]
    where \(w_1\) and \(w_2\) adjust the contributions from the message and node features, respectively.
\end{itemize}

\subsection{Loss Function}

To effectively train our model, we introduce a custom loss function that combines the mean squared error (MSE) with domain-specific penalties for predictions outside a lower limit \(L\).

The loss function is defined as:

\begin{equation}
\mathcal{L} = \frac{1}{n} \sum_{i=1}^{n} \left[ (y_i - \hat{y}_i)^2 + 
\mathbf{1}_{\{\hat{y}_i > L\}} \left( \frac{1}{1 + e^{-3(\hat{y}_i - L)}} - 1 \right) - 
\mathbf{1}_{\{\hat{y}_i \leq L\}} \cdot \frac{1}{2} \right]
\end{equation}

where:

\begin{itemize}
    \item \((y_i - \hat{y}_i)^2\): This is the Mean Squared Error between the true \(y_i\) and the predicted value \(\hat{y}_i\).
    \item \(\mathbf{1}_{\{\hat{y}_i > L\}}\): An indicator function that applies a logistic penalty when the prediction \(\hat{y}_i\) exceeds the lower limit \(L\), ensuring smoother penalties as predictions exceed this threshold.
    \item \(\mathbf{1}_{\{\hat{y}_i \leq L\}}\): An indicator function that applies a fixed penalty of \(-\frac{1}{2}\) when \(\hat{y}_i\) is less than or equal to \(L\), discouraging predictions below this limit.
    \item \(L\): The lower threshold limit, representing a critical boundary below which predictions are penalized more harshly.
\end{itemize}

\subsection{Training}
The GNN models were trained for 50 epochs on a single P100 GPU, taking approximately 2.5 hours when 7 nodes were taken and 45 minutes when 4 nodes where taken. The training utilized the Adam optimizer with a learning rate of 0.0002 and a weight decay of 5e-4.`ReduceLROnPlateau` scheduler was employed to adjust the learning rate during training.

\section{Results and discussion}

We come up with 2 observations seeing Table 1. First that the Graph with higher node features is able to capture the information of the particle more accurately compared to the one with smaller node feature. Second is that the GNN based model gives lesser Mean Average Error(MAE) compared to the TabNet models.
\begin{table}[ht]
\centering
\small
\setlength{\tabcolsep}{5pt}
\begin{tabular}{lcccc}
\toprule
\textbf{Model} & \textbf{MAE} & \textbf{Params} & \textbf{Inf. Speed} & \textbf{Epochs to} \\
 &  &  &  & \textbf{Converge} \\
\midrule
TabNet & 0.8855 & 7,488 & 0.0193 ms & 20 \\
GNN (4-dim node feat.) & 0.8850 & 99,952 & 0.1391 ms & 47 \\
GNN (7-dim node feat.) & 0.8474 & 101,152 & 0.114 ms & 18 \\
\bottomrule
\end{tabular}
\vspace{1mm}
\caption{Comparison of models on MAE, parameter count, inference speed, and convergence.}
\label{tab:model_comparison1}
\end{table}

\begin{table}[h!]
\centering
\resizebox{\textwidth}{!}{%
\begin{tabular}{|l|c|c|c|c|}
\hline
\textbf{Model}                          & \textbf{MAE} & \textbf{MSE} & \textbf{Avg Inference Time (micro-sec)} & \textbf{No. of Parameters} \\ \hline
\textbf{TabNet}                         & 0.9607            & 2.9746            & 458.7                                & 6696                                    \\ \hline
\textbf{GNN-bendAngle}                  & 1.202931          & 3.520059          & 522.204                              & 5579                                    \\ \hline
\textbf{GNN-bendAngle2}                  & 1.215189          & 3.605358          & 530.19
& 5903                                    \\ \hline
\textbf{GNN-etaValue}                   & 1.146910          & 3.240220          & 522.204                              & 5579                                    \\ \hline
\textbf{GNN-etaValue2}                  & 0.992087          & 2.525927          & 530.19                & 5903                                  \\ \hline
\textbf{GNN-etaValue3}                  & 1.145697
& 3.276628
& 614.565                                    & 6437                                 \\ \hline
\textbf{GNN-etaValue4}                  & 1.133285
& 3.205457
& 267.509                                    & 6112                                 \\ \hline
\textbf{GNN-etaValue5}                  & 0.941620
& 2.312492
& 515.472                                    & 6545                                 \\ \hline
\textbf{GNN-etaValue6}                  & 0.946957
& 2.286622
& 283.216                                    & 14176                                 \\ \hline
\end{tabular}%
}
\caption{Comparison of TabNet and GNN Models}

\label{tab:model_comparison}
\end{table}

\section{Conclusion}
In summary, our study demonstrates the effectiveness of Graph Neural Networks (GNNs) in analyzing data from CMS trigger stations. The GNN model, which captures both local and global graph structures, outperformed traditional methods like TabNet, indicating its potential for more accurate and insightful analysis in high-energy physics experiments.

\section{Acknowledgement}

\section{Broader Impact}
The use of GNN models for estimating momentum in CMS can give us a better efficiency of the trigger systems. It opens us a new domain to explore and achieve higher efficiency of the triggers and also understand high energy physics in a better way.


\appendix


\bibliographystyle{plainnat.bst}
\bibliography{references}

\begin{thebibliography}{12}
\providecommand{\natexlab}[1]{#1}
\providecommand{\url}[1]{\texttt{#1}}
\expandafter\ifx\csname urlstyle\endcsname\relax
  \providecommand{\doi}[1]{doi: #1}\else
  \providecommand{\doi}{doi: \begingroup \urlstyle{rm}\Url}\fi

\bibitem[Acosta et~al.(2018)Acosta, Brinkerhoff, Busch, Carnes, Furic, Gleyzer, Kotov, Low, Madorsky, Rorie, Scurlock, Shi, and on~behalf of~the CMS Collaboration]{Acosta_2018}
Darin Acosta, Andrew Brinkerhoff, Elena Busch, Andrew Carnes, Ivan Furic, Sergei Gleyzer, Khristian Kotov, Jia~Fu Low, Alexander Madorsky, Jamal Rorie, Bobby Scurlock, Wei Shi, and on~behalf of~the CMS Collaboration.
\newblock Boosted decision trees in the level-1 muon endcap trigger at cms.
\newblock \emph{Journal of Physics: Conference Series}, 1085\penalty0 (4):\penalty0 042042, sep 2018.
\newblock \doi{10.1088/1742-6596/1085/4/042042}.
\newblock URL \url{https://dx.doi.org/10.1088/1742-6596/1085/4/042042}.

\bibitem[Arik and Pfister(2021)]{arik2021tabnet}
Sercan~{\"O} Arik and Tomas Pfister.
\newblock Tabnet: Attentive interpretable tabular learning.
\newblock In \emph{Proceedings of the AAAI conference on artificial intelligence}, volume~35, pages 6679--6687, 2021.

\bibitem[Atwood and Towsley(2016)]{atwood2016diffusion}
James Atwood and Don Towsley.
\newblock Diffusion-convolutional neural networks.
\newblock \emph{Advances in neural information processing systems}, 29, 2016.

\bibitem[Bridle(1989)]{bridle1989training}
John Bridle.
\newblock Training stochastic model recognition algorithms as networks can lead to maximum mutual information estimation of parameters.
\newblock \emph{Advances in neural information processing systems}, 2, 1989.

\bibitem[Bruna et~al.(2013)Bruna, Zaremba, Szlam, and LeCun]{bruna2013spectral}
Joan Bruna, Wojciech Zaremba, Arthur Szlam, and Yann LeCun.
\newblock Spectral networks and locally connected networks on graphs.
\newblock \emph{arXiv preprint arXiv:1312.6203}, 2013.

\bibitem[De~Ryck et~al.(2021)De~Ryck, Lanthaler, and Mishra]{De_Ryck_2021}
Tim De~Ryck, Samuel Lanthaler, and Siddhartha Mishra.
\newblock On the approximation of functions by tanh neural networks.
\newblock \emph{Neural Networks}, 143:\penalty0 732–750, November 2021.
\newblock ISSN 0893-6080.
\newblock \doi{10.1016/j.neunet.2021.08.015}.
\newblock URL \url{http://dx.doi.org/10.1016/j.neunet.2021.08.015}.

\bibitem[Gori et~al.(2005{\natexlab{a}})Gori, Monfardini, and Scarselli]{1555942}
M.~Gori, G.~Monfardini, and F.~Scarselli.
\newblock A new model for learning in graph domains.
\newblock In \emph{Proceedings. 2005 IEEE International Joint Conference on Neural Networks, 2005.}, volume~2, pages 729--734 vol. 2, 2005{\natexlab{a}}.
\newblock \doi{10.1109/IJCNN.2005.1555942}.

\bibitem[Gori et~al.(2005{\natexlab{b}})Gori, Monfardini, and Scarselli]{gori2005new}
Marco Gori, Gabriele Monfardini, and Franco Scarselli.
\newblock A new model for learning in graph domains.
\newblock In \emph{Proceedings. 2005 IEEE international joint conference on neural networks, 2005.}, volume~2, pages 729--734. IEEE, 2005{\natexlab{b}}.

\bibitem[Micheli(2009)]{micheli2009neural}
Alessio Micheli.
\newblock Neural network for graphs: A contextual constructive approach.
\newblock \emph{IEEE Transactions on Neural Networks}, 20\penalty0 (3):\penalty0 498--511, 2009.

\bibitem[Nair and Hinton(2010)]{nair2010rectified}
Vinod Nair and Geoffrey~E Hinton.
\newblock Rectified linear units improve restricted boltzmann machines.
\newblock In \emph{Proceedings of the 27th international conference on machine learning (ICML-10)}, pages 807--814, 2010.

\bibitem[Rumelhart et~al.(1986)Rumelhart, Hinton, and Williams]{rumelhart1986learning}
David~E Rumelhart, Geoffrey~E Hinton, and Ronald~J Williams.
\newblock Learning representations by back-propagating errors.
\newblock \emph{nature}, 323\penalty0 (6088):\penalty0 533--536, 1986.

\bibitem[Scarselli et~al.(2008)Scarselli, Gori, Tsoi, Hagenbuchner, and Monfardini]{scarselli2008graph}
Franco Scarselli, Marco Gori, Ah~Chung Tsoi, Markus Hagenbuchner, and Gabriele Monfardini.
\newblock The graph neural network model.
\newblock \emph{IEEE transactions on neural networks}, 20\penalty0 (1):\penalty0 61--80, 2008.

\end{thebibliography}
\end{document}